\documentclass[11pt]{article}
\usepackage{moriond,epsfig}

\def\be{\begin{equation}}
\def\ee{\end{equation}}
\def\bea{\begin{eqnarray}}
\def\eea{\end{eqnarray}}

\begin{document}
\vspace*{4cm}
\title{The GIM Mechanism: origin, predictions and recent uses\footnote{Opening Talk, Rencontres de Moriond, EW Interactions and Unified Theories, La Thuile, Valle d'Aosta,  Italia, 2-9 March, 2013.}
}

\author{ Luciano Maiani }

\address{CERN, Geneva, Switzerland \\
Dipartimento di Fisica, Universit\'a di Roma {\it La Sapienza, Roma, Italy}}

\maketitle\abstracts{The GIM Mechanism was introduced by Sheldon L.  Glashow, John Iliopoulos and Luciano Maiani in 1970, to explain the suppression of Delta S=1, 2 neutral current processes and is an important element of the unified theories of the weak and electromagnetic interactions. Origin, predictions and uses of the GIM Mechanism are illustrated. Flavor changing neutral current processes (FCNC) represent today an important benchmark for the Standard Theory and give strong limitations to theories that go beyond ST in the few TeV region. Ideas on the ways constraints on FCNC may be imposed are briefly described.
}

\section{Setting the scene}

The final years of the sixties marked an important transition period for the physics of elementary particles. Two important discoveries made in the early sixties, quarks and Cabibbo theory, had been consolidated by a wealth of new data and there was the feeling that a more complete picture of the strong and of the electro-weak interactions could be at hand. 

The quark model of hadrons introduced by M. Gell-Mann \cite{GellMann:1964nj} and G. Zweig \cite{Zweig:1964jf} explained neatly the spectrum of the lowest lying mesons and baryons as $q\bar q$ and $qqq$ states, respectively, with:
\bea
&& q=\left[ \begin{array}{c} u\\ d\\s \end{array}\right]
\label{simplequark}
\eea
and electric charges $Q=2/3, -1/3, -1/3$, respectively.

Doubts persisted on quarks being real physical entities or rather a simple mathematical tool to describe hadrons
. This was due to the {\it symmetry puzzle}, namely the fact that one had to assume an overall symmetric configuration of the three quarks to describe the spin-charge structure of the baryons, rather than the antisymmetric configuration required by the fermion nature of physical, spin $1/2$, quarks\footnote{The problem is exemplified by the spin $3/2$ resonance, $\Delta^{++}$ . With the quark structure in (\ref{simplequark}), the state $\Delta^{++}$ in the spin state with $S_z=3/2$ has to be described as $\Delta^{++}(S_z=3/2)=[u^\uparrow u^\uparrow u^\uparrow]$which is an S-wave, symmetric state.}. Ignoring these doubts, there had been already attempts to describe the strong interactions among quarks with the exchange of a {\it neutral vector particle}, which had even been given a specific name,  the {\it gluon}, derived  from its role  to {\it glue} the quarks inside the hadrons.

It has to be said that the efforts to construct a fundamental strong interaction theory were focussed, in these years, rather along the lines indicated by the pioneering paper by G. Veneziano \cite{veneziano1}. However, the resulting {\it dual models} \cite{56697} later evolved in the Nambu \cite{EFI 74/40}-Goto \cite{68024} {\it string theory} are entirely outside the scope of the present review.

The Cabibbo theory \cite{weakangle} had extended to the weak decays of strange particles the idea of universality pioneered by E. Fermi and later developed by R. Marshak and C. G. Sudarshan \cite{MarSud},  R. P. Feynman and M. Gell-Mann \cite{FGM1}, S. S. Gerstein and Ya. B. Zeldovich \cite{GZEL},  and others.

The Weak Interaction lagrangian was 
described by the product of a weak current, $J^\lambda$, with its hermitian conjugate, currents being, in turn, made of universal pieces representing lepton and hadron contributions. In formulae:
\begin{eqnarray}
&&J^\lambda = {\bar \nu}_e \gamma^\lambda(1-\gamma_5) e +{\bar \nu}_\mu \gamma^\lambda(1-\gamma_5) \mu + {\bar u} \gamma^\lambda(1-\gamma_5) d_C\label{weakcurr} \\ 
&&d_C = \cos\theta d + \sin\theta s\label{cabibbo}\\
&&{\cal L}_F = \frac{G_F}{\sqrt{2}}J^\lambda J_\lambda^+
\label{fermilagr}
\end{eqnarray}
with $G_F$ the Fermi constant taken from muon decay and $\theta$ the Cabibbo angle.

The  ultraviolet divergent character of the Fermi interaction invited to think that the lagrangian (\ref{fermilagr}) could be the effective lagrangian of a less divergent or even renormalizable theory. Two options were conceived at the time. The simplest possibility (the Intermediate Vector Boson hypothesis, IVB) was to assume the current $\times$ current interaction to be mediated by a single, electrically charged, massive vector boson, $W$, with:  
\begin{eqnarray}
&&{\cal L}_W =g W_\lambda J^\lambda+{\rm h.c.} \nonumber \\
&& \frac{G_F}{\sqrt{2}} = \frac{g^2}{M_W^2}
\end{eqnarray} 
	
	The second possibility, first considered by J. Schwinger, was to imbed the IVB theory into a fully fledged Yang-Mills theory, with the interaction determined by local invariance under a non-abelian gauge group, which would eventually include the electromagnetic gauge invariance\footnote{Schwinger \cite{schwinger} put the electromagnetic with the weak interactions in the simple group $SU(2)$; S. Bludman \cite{blud} formulated the first SU(2) gauge theory of weak interactions alone.}. A theory of the electro-weak interactions of leptons based on the gauge group $SU(2)\otimes U(1)$ had been proposed by S. Glashow in 1961 \cite{Glashow:1961tr}. 
	
	Glashow's theory predicted two neutral vectors, the massless photon and a massive neutral boson, called $Z^0$, in addition to the charged IVB. The breaking of the gauge group from $SU(2)\otimes U(1)$ down to the electromagnetic gauge group $U(1)_Q$ was enforced by the explicit addition of mass terms for the vector bosons and for the leptons. It was hoped that such a breaking, associated to operators of dimension less than four, would not disturb the renormalizability of the Yang-Mills theory. 
	
	Two results, proven in the late sixties are here relevant. First, the renormalizability of the massless (i.e. exactly gauge symmetric) Yang-Mills theory, by E. S. Fradkin and I. V. Tyutin \cite{frad&tyu}. Second, the investigations by M. Veltman \cite{veltmanetal} which stressed the singular character of the vector boson mass term and the consequent fact that the theory with explicit mass terms was probably non-renormalizable.
	
	In their seminal papers, S. Weinberg \cite{Weinberg:1967tq} and A. Salam \cite{salam} proposed a new way to a realistic, unified electroweak theory. This was based on the spontaneous symmetry breaking of the gauge symmetry induced by a non-vanishing vacuum expectation value of a scalar field. The newly discovered formulation of the spontaneous breaking of a gauge symmetry, by P. Higgs \cite{Higgs:1966ev} and by F. Englert and R. Brout \cite{englert}, implied that the massless Goldstone bosons, associated to the spontaneous symmetry breaking of a continuous, global symmetry, were absorbed into the longitudinal degrees of freedom of  {\it massive vector bosons}. The  $SU(2)\otimes U(1)$ gauge theory was reformulated in this direction, with the hope that the exact gauge symmetry of the equations of motion could produce a renormalizable theory, an hypothesis that was to be proven correct by G. 't-Hooft and M. Veltman \cite{'tHooft:1972fi} a few years later.
	
	Similarly to Glashow's 1961, the theory of Weinberg and Salam could be applied only to the electroweak interactions of the leptons. It was immediate to imbed the Cabibbo hadron current into an $SU(2)\otimes U(1)$ algebra, but it was also immediate to see that the neutral current coupled to the $Z^0$ would carry a violation of strangeness of a size already excluded by the non observation, at the time, of the neutral current weak decay:
\be
K_L \to \mu^+ \mu^-
\label{kltomumu} 
\ee

To see this, it is enough to rewrite the hadron current in (\ref{weakcurr}) in matrix form, according to:
	
\begin{eqnarray} 
&&J^{had}_\lambda ={\bar u} \gamma_\lambda(1-\gamma_5) d_C={\bar q} \gamma_\lambda(1-\gamma_5) {\cal C}q \nonumber \\
&& \nonumber \\
&& {\cal C} =\left( \begin{array}{ccc} 0 & \cos\theta &\sin\theta \\ 0 & 0 & 0 \\ 0 & 0 & 0 \end{array}\right)
\label{matrixcab}
\end{eqnarray} 

Writing: 
\be
{\cal C}=(I_W)_1+i(I_W)_2; \;\;{\cal C}^\dagger= (I_W)_1-i(I_W)_2
\ee
one sees ${\cal C}$ and ${\cal C}^\dagger$ as the raising and lowering elements of a weak-$SU(2)$ algebra whose third generator is given by:
\be
2(I_W)_3 = \left[{\cal C}, {\cal C}^\dagger \right] = \left[ \begin{array}{ccc} 1 &0 & 0 \\ 0 & -\cos^2\theta & -\cos \theta \sin \theta \\ 0 & -\cos \theta \sin \theta &-\sin^2\theta \end{array}\right]
\label{thirdcomp}
\ee

In the Weiberg-Salam and Glashow theories, the physical vector boson, $Z^0$, is coupled to a linear combination of the electromagnetic current (which is flavour diagonal) and of the current associated to $(I_W)_3$. One is then left with a Flavour  Changing Neutral Current (FCNC) determined by the Cabibbo angle and totally excluded by the data. For reference, the present value of the ratio of the leptonic FCNC rate of $K_L$ to the leptonic rate of $K^+$ is \cite{PDG}:
\be
\frac{\Gamma(K_L \to \mu^+ \mu^-)}{\Gamma(K^+ \to \mu^+ {\bar \nu}_\mu)}= 2.60\times 10^{-9}
\ee
\section{The Joffe-Shabalin cutoff}

In mid 1968, a paper by B. L. Ioffe and E. P. Shabalin \cite{i&s} shed a sudden light on the structure of higher order weak interactions. The paper addressed the calculation of amplitudes with the exchange of two IVBs, considering several FCNC amplitudes like the weak amplitude of (\ref{kltomumu}), $\Delta S\pm 1$, and the $K^0 \to \bar K^0$ mixing amplitude, $\Delta S=2$. Similar results were obtained by F. Low~\cite{flow} and by R. Marshak and collaborators \cite{marshaketal}.

Contrary to the belief that strong interaction form factors would soften all divergences of the IVB theory, the authors found quadratically divergent  amplitudes, which therefore had to be regulated by an ultraviolet cutoff, $\Lambda$. The singular behaviour was derived from the equal time commutators of the weak currents, which indicated that hadrons are indeed soft objects, but are made of point-like constituents. In the same year, experiments at SLAC showed a point-like behaviour of the deep-inelastic cross section of electrons on protons and deuterons, giving the first experimental indication of point like  constituents inside the nucleons.

Most surprisingly, the large suppression of FCNC amplitudes required a small value of the cutoff. The more stringent limit is given by the $K^0 \to \bar K^0$ transition, which requires $\Lambda_{I\&S}\simeq 3$ GeV.

The calculation of ref.~\cite{i&s} addressed FCNC amplitudes of order $G(G\Lambda^2)$. It was also realised that quadratic divergent amplitudes would arise, in the IVB theory, to order $G\Lambda^2$, potentially providing violations to the symmetries of the strong interaction amplitudes (parity, isospin, $SU(3)$ and strangeness). C. Bouchiat, J. Iliopoulos and J. Prentki \cite{bip} showed that, with $SU(3)\otimes SU(3)$ breaking described by a $(3, {\bar 3})$ representation, the leading divergences give only diagonal contributions, hence no parity and strangeness violations.

Attempts were made in 1968-1969 to cope with the disturbing divergence by:
\begin{itemize} 
 \item 
introducing spin zero and spin one intermediaries with compensating couplings so as to move the divergence into diagonal, flavour conserving, amplitudes \cite{gmlkr} (it was soon understood that too many were needed);
\item 
cancelling the isospin violation effect due to the $G\Lambda^2$ divergences with a specific value of the Cabibbo angle \cite{gattoetal} \cite{nclm}, related to the parameters of the $(3, {\bar 3})$ symmetry breaking (quark masses, in modern terms);
\item
introducing negative metric states \cite{leewick}.
\end{itemize}  

\section{The GIM mechanism}

In modern terms, the result of Ioffe and Shabalin for the  $K^0 \to \bar K^0$ amplitude derives from the Feynman diagrams reported in Fig. 1, restricted to $u$-exchange diagrams. 
Weak couplings at the quark vertices are indicated, the most divergent part of the amplitude is:
\be
A^{u-exch}(K^0 \to \bar K^0)= Const \; G(G\Lambda^2) \sin^2\theta \cos^2 \theta
\label{divergent} 
\ee

\begin{figure}[htb]
\begin{center}
\includegraphics[scale=.50]{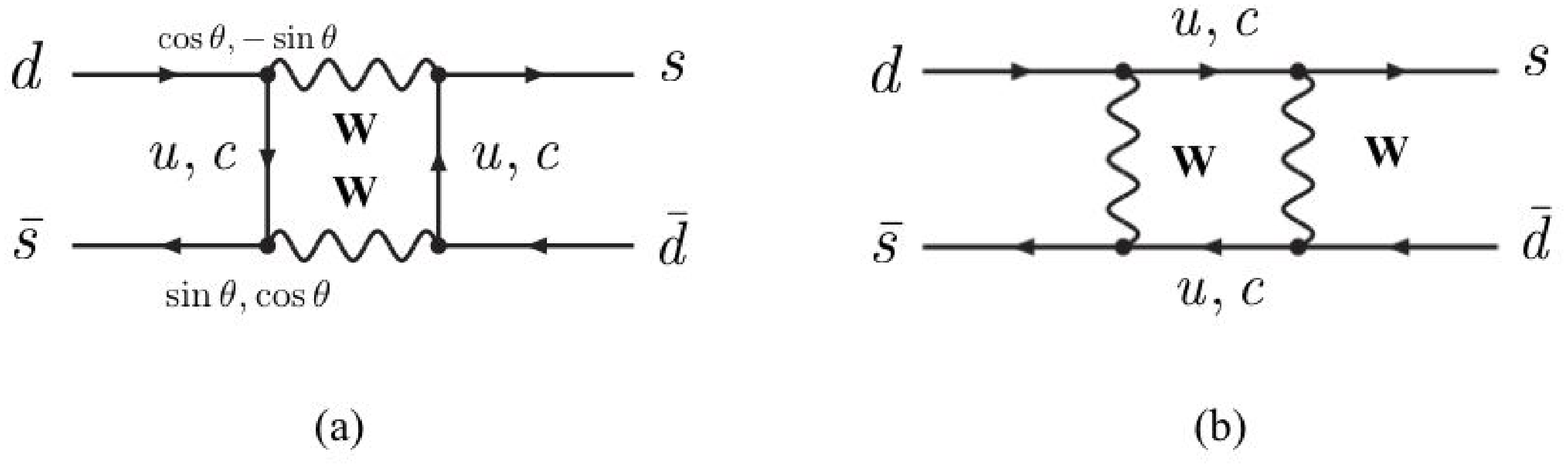}
\caption{
{\small 
Quark diagrams for  $K_L \to \mu^+ \mu^-$ after GIM. Couplings of $u$ and $c$ to $d$ and $s$ quarks are indicated.} 
}
\label{univ}
\end{center}
\end{figure}

The solution proposed by Glashow, Iliopoulos and Maiani \cite{GIM70} in 1970 requires the existence of a new, charge $2/3$ quark, the charm quark, coupled by the weak interaction to the  superposition of $d$ and $s$ quarks {\it orthogonal} to the  Cabibbo combination $d_C$, eq. (\ref{cabibbo}). The new term in the weak current is:
\begin{eqnarray} 
&&J_\lambda^{(charm)}=\bar c \gamma_\lambda(1-\gamma_5)s_C; \nonumber  \\
&&s_C= -\sin\theta d + \cos\theta s
\label{gimcurrent}
\end{eqnarray} 

For each $u$-line exchanged, the charm quark provides a second diagram with a coupling of opposite sign. 
In fact, were the mass of the charmed quark equal to the mass of the up quark, the two diagrams would exactly cancel. For unequal masses, the result  must be proportional to the difference $m_c^2-m_u^2$. It is easy to see that the quark mass-squared difference takes the role of the ultraviolet cutoff in (\ref{divergent}). The Ioffe and Shabalin estimate of $\Lambda$ turns into a prediction of the charm quark mass (neglecting $m_u$):
\be
 m_c \simeq \Lambda_{I\&S} \simeq 3\;{\rm GeV}
\label{cmass}
\ee 

Similar considerations apply to the $K_L \to \mu^+ \mu^-$ transition, which however leads to a less stringent prediction of the cutoff, due to the fact that the transition also proceeds also via the electromagnetic, $\gamma-\gamma$ intermediate state.

The diagrams in Fig. 1 and the prediction (\ref{cmass}) describe in essence the GIM mechanism.


\section{Electroweak unification with the charm quark}  

Besides adding a new chapter to hadron spectroscopy, the extension to the charm quark gave a decisive momentum to the unification of electromagnetic and weak interactions. With the GIM addition, the weak interaction matrix $\cal C$, eq. (\ref{matrixcab}), takes the form (we order the four quarks as $u, c, d, s$) :
\be
\cal C= \left(\begin{array}{cc} 0 & U \\ 0 & 0 \end{array}\right)
\label{gimneutralcurr}  
\ee 
where $U$ is  two by two and orthogonal:
\be
U= \left(\begin{array}{cc}\cos\theta & \sin\theta \\ -\sin \theta & \cos \theta \end{array}\right)
\label{gimmatrix} 
\ee

The orthogonality of $U$ makes so that the neutral current associated to the third component of the weak isospin is flavour diagonal:
\be
2(I_W)_3 = \left[{\cal C}, {\cal C}^\dagger \right] = \left(\begin{array}{cc} U U^T&0 \\ 0&-U^T U \end{array}\right)= \left( \begin{array}{cc} 1&0 \\ 0&-1\end{array}\right)
\label{ithreegim}
\ee

With the charm quark, a Yang-Mills electroweak theory based on $SU(2) \otimes U(1)$ becomes possible for quark and leptons. FCNC processes are forbidden at tree level by the structure of $(I_W)_3$ in eq. (\ref{ithreegim}). They are possibile in higher order, by the diagrams reported in Fig. 2 for the typical case of $K^0 \to \mu^+ \mu^-$, but suppressed by the GIM mechanism. The not-too-large value of the charm mass makes these amplitudes to be effectively of order $G^2$, as required by observation. The analysis of FCNC in the full gauge theory, performed by M. K. Gaillard and B. W. Lee \cite{mk-bwl} has later confirmed the general order of magnitude of $m_c$ in (\ref{cmass}). 
\begin{figure}[htb]
\begin{center}
\includegraphics[scale=.40]{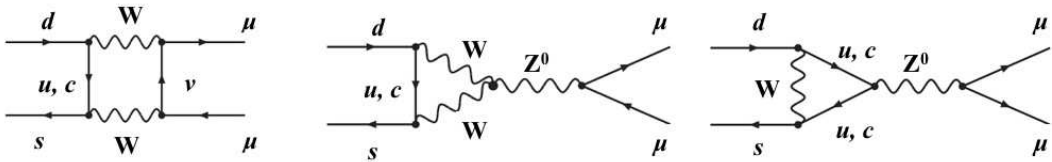}
\caption{
{\small 
Feynman diagrams for: $K^0=(d{\bar s})\to \mu^+\mu^-$ in the electroweak gauge theory. Higgs boson exchange diagrams to be added.} 
}
\label{univ}
\end{center}
\end{figure}
\paragraph{Neutrino neutral current processes.}

With the structure in eqs. (\ref{gimneutralcurr}) and (\ref{gimmatrix}), flavour conserving, neutral current processes are indeed predicted to occur with similar rates as charged current processes. Neutrino scattering off nucleons leading to a final state without muons are considered in ref. \cite{GIM70}, with reference to the neutral current of a pure Yang-Mills theory. Cross sections were obtained for the inelastic processes:
\be
\nu (\bar \nu) + {\rm Nucleous} \to \nu (\bar \nu) + {\rm hadrons}
\label{muonless}
\ee
 not much below the existing limits of the time, thus indicating neutrino neutral current processes to be a promising signal for the proposed four quark theory. 
 
 In 1973, the Gargamelle bubble chamber collaboration at CERN observed what they called {\it muonless } or {\it electronless} neutrino events \cite{gargamelle73}, i.e. multihadron neutrino interactions without a visible muon or electron track in the final state, soon recognised to be neutrino processes of the type (\ref{muonless}). Detailed analysis showed that indeed strange particles are pair produced in the final state, indicating flavour conservation in these abundant neutral current reactions.

\paragraph{Quark-lepton symmetry.}

Quark-lepton symmetry is not respected in the weak current (\ref{weakcurr}), which features two lepton isospin doublets and only one quark doublet. Restoring quark-lepton symmetry was one of the basic motivations of the GIM paper \cite{GIM70}, providing the basis for the partial cancellation of FCNC amplitudes. 

It is worth noticing that quark-lepton symmetry plays amore  fundamental role in the unified electroweak theory. C. Bouchiat, J. Iliopoulos and P. Meyer \cite{BIM72} have shown that the symmetry is the basis for the cancellation of the Adler-Bell-Jackiw anomalies, the last obstacle towards a renormalizable theory,  for fractionally charged and $SU(3)_{color}$ triplet quarks.

\paragraph{CP violation, in brief.}
It was recognized in  \cite{GIM70} that the, generally complex, matrix $U$ arises from the diagonalization of the quark mass matrix. It was also noted there that with four quarks one can always bring $U$ into the real form  (\ref{gimmatrix}), thereby excluding CP violation from the weak interaction. Already worried by the charm quark, we did not ask what would happen with even more quarks and failed to discover a simple theory of CP violation.

 Three years later, Kobayashi and Maskawa \cite{KM73} showed that a complex phase does remain if the matrix (now currently indicated as  $U_{CKM}$, after Cabibbo, Kobayashi and Maskawa)  is three by three, making it possible to incorporate the observed CP violation in a theory with six quark flavours. 

The phenomenology of CP violation with six quarks has been first explored by S. Pakvasa and H. Sugawara \cite{Pakvasa:1975ti} and by L. Maiani \cite{Maiani:1975in}. 

In 1986, I. Bigi and A. Sanda predicted direct CP violation in B decay; in 2001, Belle and BaBar discover CP violating mixing effects in B-decays.

\begin{figure}[ht]
\begin{center}
\includegraphics[scale=.35]{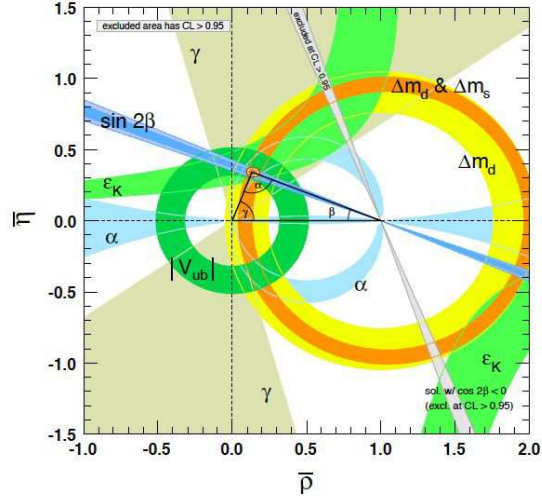}
\caption{{\small CKM fit. The vertex of the unitarity triangle gives the values of  $\rho$ and $\eta$ as determined from the intersection of the regions determined by the different observables, $\epsilon_K, \Delta m$, etc.
.}}
\label{CKMfit}
\end{center}
\end{figure}

Today, the description provided by the U$_{CKM}$ matrix has met with an extraordinary success. In Wolfenstein's parametrization \cite{wolfckm}:

\bea
{\small 
U_{CKM} =\left(\begin{array} {ccc} 1-\frac{1}{2}\lambda^2
& \lambda &A\lambda^3(\rho-i\eta)\\
  &   &   \\
-\lambda & 1-\frac{1}{2}\lambda^2 & A\lambda^2 \\
 &   &   \\
A\lambda^3[1- (\rho+i\eta)] & -A\lambda^2 
& 1 \end{array}\right) 
}
\eea
Fig. \ref{CKMfit} illustrates the excellent fit obtained for the parameters $\rho$ and $\eta$ from the measurements of different observables in $K$ and $B$ physics  \cite{Ceccucci:2008zz}.  

\section{Precursors and discovery of charmed particles}
In mid nineteen fifties, the Sakata model \cite{sakata1} featured three basic constituents of the hadrons, (p, n, $\Lambda$), in parallel to the three elementary leptons known at the time:
\bea
{\rm elementary \;hadrons} =\;\left(\begin{array}{ccc} & p &   \\ 
n &  & \Lambda\end{array}\right);\;
\;\; {\rm leptons}=\;\left(\begin{array}{ccc} & \nu &  \\e &  & \mu\end{array}\right)
\eea

In 1962, after the discovery of the muonic neutrino, Sakata and collaborators \cite{sakata2} , at Nagoya,  and Katayama and collaborators  \cite{katayama}, in Tokyo, proposed to extend the model to a fourth baryon,  called V$^+$:
\bea 
{\rm elementary \;hadrons} =\left(\begin{array}{cc} p &V^+  \\  n & \Lambda\end{array}\right);\;
\;\; {\rm leptons}=\left(\begin{array}{cc}\nu_1 & \nu_2   \\ e & \mu \end{array}\right)
\eea
a possible mixing among  $\nu_e$ and $\nu_\mu$ was paralleled by $n-\Lambda$ mixing {\it \' a-la} Cabibbo, giving rise to weak couplings of p and $V^+$ similar to the ones we have assumed for $u$ and $c$.  


Restoring quark-lepton symmetry had also been the reason for the early consideration of the fourth quark, $c$, by J. Bjorken and S. Glashow \cite{BJ&G}, where the weak coupling (\ref{gimcurrent}) was also written explicitly. 

The lack of any connection to the FCNC processes and, consequently, the lack of  information on the mass-scale of the charm quark, prevented further progress. 

Indeed, in 1970, there was no experimental evidence of weakly decaying hadrons beyond the lowest lying strange baryons and mesons. The fact that hadrons could be made with only three types of quarks was accepted as almost self evident. The theory of charm had to explain first of all why,  accelerator energies being already well above the mass scale indicated by (\ref{cmass}), none of the charm particles had been seen. This question was answered in \cite{GIM70}. 

{\it Why have none of these charmed particles been seen? Suppose they are all relatively heavy, say 2 GeV. Although some of the states must be stable under strong (charm-conserving) interactions, these will decay rapidly ( $10^{-13}$sec) by weak interactions into a very wide variety of uncharmed final states (there are about a hundred distinct decay channels). Since the charmed particles are copiously produced only in associated production, such events will necessarily be of very complex topology, involving the plentiful decay products of both charmed states. Charmed particles could easily have escaped notice.}

In fact, starting from 1971, emulsions experiments performed in Japan by K. Niu and collaborators  \cite{Niu_charm}, did show cosmic ray events with {\it kinks}, indicating long lived particles (on the nuclear interaction time scales) with lifetimes in the order of $10^{-12}$ to $10^{-13}$  sec. These lifetimes are in the right ballpark for charmed particles and indeed they were identified as such in Japan (see again \cite{Niu_charm} for a more details). However, cosmic rays events were paid not much attention in western countries.

The first unequivocal evidence for a $c\bar c$ state was provided  in 1974 by the $J/\Psi$ particle ($M_{J/\Psi}=3.097$ GeV) discovered by C. C. Ting and collaborators \cite{J_ting} at Brookhaven, by B. Richter and collaborators \cite{J_richter} at SLAC and immediately after observed in Frascati \cite{LNF-74/61-P}. The discovery came with the surprise that the $J/\Psi$ was much narrower than anticipated in  \cite{GIM70}. This was interpreted \cite{drgg} as a manifestation of the {\it asymptotic freedom} at small distances of the color forces, which bind the quark-antiquark pair, recently discovered by D. Gross and F. Wilczek \cite{QCD1} and by D. Politzer \cite{QCD2}. 
The charm quark is heavy enough for the $c\bar c$ pair to be separated by distances small enough for color forces to be already in the small coupling regime.  

The color quark degrees of freedom solved at once the symmetry puzzle mentioned in Sect. 1 and the massless color quanta replaced the single gluon introduced in the early sxties.

{\it Naked charm particles} have been searched in $e^+ e^-$ colliders by the most visible signature, the so-called $e-\mu$ events, originated by the semileptonic decays of the lightest charmed particles:
\bea
&& e^+ e^- \to c+\bar c \to e^+ ({\rm or} \; \mu^+) + \mu^- ({\rm or} \;e^-) + {\rm hadrons}
\eea

Events of this kind were in fact observed by M. Perl and collaborators \cite{tau_perl}  at energies above the $J/\Psi$, but with the wrong energy distribution of the leptons. A state of confusion ensued, until it was realized that pairs were being produced of an entirely unexpected new particle, which also decayed semileptonically. This was the heavy lepton $\tau$, whose threshold, for yet unexplained reasons, happens to coincide quite precisely with the $c\bar c$ threshold. 

It was only in 1976 that this fact was clearly recognised. The multihadron events in $e^+ e^-$ annihilation, depured of $\tau$-pairs events, showed clearly the  $c\bar c$ threshold with the jump in the cross section of the size corresponding to a spin 1/2, charge 2/3, particle.

The lightest weakly decaying charmed meson, $D^0 = (c\bar u)$ ($M_{D^0}=1.865$ GeV) was discovered in 1976 by the Mark I detector \cite{dmeson} at the Stanford Linear Accelerator Center. The charged meson $D^+=(c\bar d)$ ($M_{D^+}=1.870$ GeV) and the lowest lying baryons, $\Lambda^+_c =[c(ud)_{I=0}]$ ($M_{\Lambda^+_c}= 2.286$ GeV), and $\Sigma^+_c =[c(ud)_{I=1}]$ ($M_{\Sigma^+_c}= 2.453$ GeV), soon followed.

The same year, L. Lederman and collaborators, studying $p \bar p$ collisions at Fermilab, observed a new narrow state \cite{b_lederman}, the $\Upsilon$ particle. The  $\Upsilon$ is similar to the $J/\Psi$ but made by a heavier quark, soon identified with a charge -1/3 quark named $b$-quark ($b$ for {\it beauty}). 

At the same time that the charm quark and charm spectroscopy were discovered with properties very close to what predicted in \cite{GIM70}, the third generation of quarks and leptons, anticipated by Kobayashi and Maskawa on the basis of the observed CP violation in K decays, was being unveiled. 

\section{FCNC in the Standard Theory}

Quark loops with d $\to$ s, b transitions, which describe $\Delta$F=2 transitions for K and B, are dominated by c and t quarks. 
The leading QCD corrections are represented by multiplicative renormalizations to loop amplitudes, reliably calculable for loops dominated by $c$ or $t$ quark, due to asymptotic freedom.
For the $K^0$ or $B^0$ system, to lowest electroweak order, one finds the general formula:
\bea
&& M_{12}(\bar K^0\to K^0) =  < K^0|-{\cal L}_{eff}|\bar K^0>=\nonumber \\
&&=\frac{(G_F M_W^2)(G_F f_K^2)}{12\pi^2}  \times \sum_{i,j=c,t}C_i C_j E(x_i, x_j)\times m_K
\label{deltam}
\eea   
where the $C_i$ and $C_j$ are combinations of CKM coefficients and $E(x_i, x_j)$ are loop factors \cite{inamilim} with $x_{c,t}=m_{c,t}^2/M_W^2$. Color corrections renormalize the various terms of (\ref{deltam}) according to \cite{burasetal,ciuchini}:
\bea
&& M_{12}(\bar K^0\to K^0)|_{corr}=\frac{(G_F M_W^2)(G_F f_K^2)}{12\pi^2}  \times \nonumber \\
&& \times \left[ \eta_1 C_c^2E(x_c, x_c)   +   \eta_2 C_t^2 E(x_t, x_t) +2 \eta_3 C_c C_t E(x_c, x_t)\right]\times m_K  \times B_K 
\eea
For D-mesons, $\Delta$F=2 transitions are dominated by $s$ and $b$ quarks. Since: 
\be
C_b\approx (\sin\theta_C)^5, \;{\rm vs.}\; C_s\approx (\sin\theta_C) 
\ee
$b$ is CKM suppressed much more than $s$ and long-distance effects dominate.

In Tab. \ref{FCNCdata} the predicted FCNC observables in $K$ and $B$ mesons thus far determined are confronted to the experimental data, including the branching ratio of $B_s^0 \to \mu^+ \mu^-$,  recently observed by the LHCb Collaboration \cite{bstomumu}.

\begin{table}[htb]
\caption{{\footnotesize FCNC effects in K and B mesons. Masses in MeV. Input values: 
$m_c=1.5$, $m_t=173$.}}
\label{confronto_dati}
\begin{center}
\vskip0.3cm
\begin{tabular}{@{}|c|c|c|c|c|c|l|}
\hline 
      & $|\epsilon_K|$  &  $\Delta m_K$  &  $|\Delta M(B^0_d)|$  &    $|\Delta M(B^0_s)|$& $Br(B_s \to \mu^+\mu^-)$ \\
  \hline     
EW diagr. & {\small $6.34\;10^{-3}$  } & {\small $3.12\;10^{-12}$ } & {\small $7.51\;10^{-10}$}  &  {\small $294\;10^{-10}$} &{\small $4.0\;10^{-9}$ } \\ 
\hline
 QCD corrcts &  {\small $2.65\;10^{-3}$ }   &  {\small $3.85\;10^{-12}$ }   &   {\small $4.13\;10^{-10}$ }  &  {\small $119\;10^{-10}$ }&{\small $(3.53\pm0.38)\;10^{-9}$ } \\
\hline
            expt  &  {\small $2.228\;10^{-3}$ }    &   {\small $3.483\;10^{-12}$ } &   {\small $3.34\;10^{-10}$ }  &  {\small $117.0\;10^{-10}$ }& {\small $(3.2\pm1.4)\;10^{-9}$ } \\
 \hline
\end{tabular}\\[2pt]
\label{FCNCdata}
\end{center}
\end{table}
Due to the values of CKM parameters, the mass difference $\Delta m_K$ is dominated by the charm quark, and the original GIM estimate is reproduced even with three quark generations. The other observables are dominated by the $t$ quark. The agreement between the values in the last two lines is indeed remarkable.

\section{Recent uses of GIM}

There is a widespread belief that the Standard Electroweak Theory must be completed into some high energy theory. One argument is that the theory does not contain  gravity and/or Grand Unification of particles interactions. One expects modifications when quantum gravity effects become relevant, that is at the Planck mass, $M_ {Planck}\simeq 10^ {19}\;$GeV, or at the Grand Unification mass, $M_{GUT}\simeq 10^ {16}\;$GeV. 

Arguments that the energy for the new regime may start at much smaller energies, of the order of 1 TeV, come from the so called  {\it unnaturalness} of the Standard Theory \cite {hooft} in connection with the value of the Higgs boson mass, and from the {\it hierarchy problem} \cite{hierarchy}. 

Quantum corrections to the Higgs boson mass, computed at one loop with some ultraviolet cutoff, $\Lambda$, are quadratically divergent:
\be
\mu^2 = \mu_0^2 +Const\; \frac{\alpha}{\pi}\;  \Lambda^2
\label{quantumcorrr} 
\ee
where $\alpha$ is the fine structure constant. 

The conspiracy between $\mu_0^2$ and $\Lambda^2$ to produce the physical value of the mass becomes more and more unnatural at the increase of $\Lambda$. A natural situation would require:
\be
 \mu^2 \approx \frac{\alpha}{\pi} \times \Lambda^2
\ee

For $\mu = 125$ GeV, the mass of the particle observed by ATLAS and CMS \cite{125}, this leads to:
\be
\Lambda \approx 2\;{\rm TeV}
\ee
A light Higgs boson requires new physics in the TeV range.

Eq. (\ref{quantumcorrr}) can be contrasted with the correction to the mass of a fermion, which is of the form:
\be
m= m_0 + Const\;\frac{\alpha}{\pi} \;m_0 \log (\frac{\Lambda^2}{Q^2})
\ee
The correction must be proportional to $m_0$ because a new symmetry is gained for $m_0=0$, chiral symmetry, which requires the physical mass to vanish as well. Thus the correction diverges only logarithmically with $\Lambda$, which can be as large as $M_ {Planck}$ or $M_ {GUT}$. In the Standard theory, no new symmetry is recovered when a scalar particle mass vanishes, which makes so that the Higgs boson mass is naturally of the order of the cutoff.
\paragraph{Predicting supersymmetry ?}  Quantum corrections to $\mu_0^2$ are made of loop with bosons and fermions, entering with opposite sign. Thus cancellations are possible, to make the result more convergent. This is the case if the Standard Theory is embedded into a larger theory with supersymmetry \cite{susy}. Supersymmetry protects the scalar particle mass by relating it to the fermion mass. Were supersymmetry exact, the quadratic correction to  $\mu_0^2$ would vanish exactly. For broken supersymmetry, the role of the cutoff in (\ref{quantumcorrr}) is replaced \cite{susypred} by the mass-squared of the lightest supersymmetric partners of the vector bosons, leptons and quarks:
\be
M_{SUSY} \simeq \Lambda = {\cal O}(1{\rm TeV})
\label{msusy}
\ee

The similarity with the GIM argument, eq. (\ref{cmass}),  is evident.
 


Let's suppose then  that, in line with the above arguments, a light Brout-Englert-Higgs scalar boson requires new physics (NP), at energy scale of the order of the TeV scale, be it SUSY, or new composite states, or other.

The new particles most likely will carry flavor and will potentially add new FCNC effects to the pattern predicted by the Standard Theory alone.

To analyze the situation, one may assume that the new physics will produce, at low energy, additional effects described by non-renormalizable, dimension six, operators added to the effective lagrangian of the Standard Theory. One may write, for example:
\bea
&&{\cal L}_{eff}(d \bar s\to \bar d s)= \nonumber \\
&&= -\frac{G_F^2 M_W^2}{16\pi^2}  \times \sum_{i,j=c,t}(U^*_{id}U_{is})( U^*_{jd}U_{js}) E(x_i, x_j)\times
\left(\bar d s\right)_{V-A}\left(\bar d s\right)_{V-A}+  \nonumber \\
&&+ \sum_i \frac{(c_{NP})_i}{\Lambda^2}\; {\cal O}_i
\label{np_efflagr}
\eea
with a suitable set of operators ${\cal O}_i$ and with a cutoff $\Lambda$ characterizing the NP energy scale.

\begin{figure}[ht]
\begin{center}
\includegraphics[scale=.5]{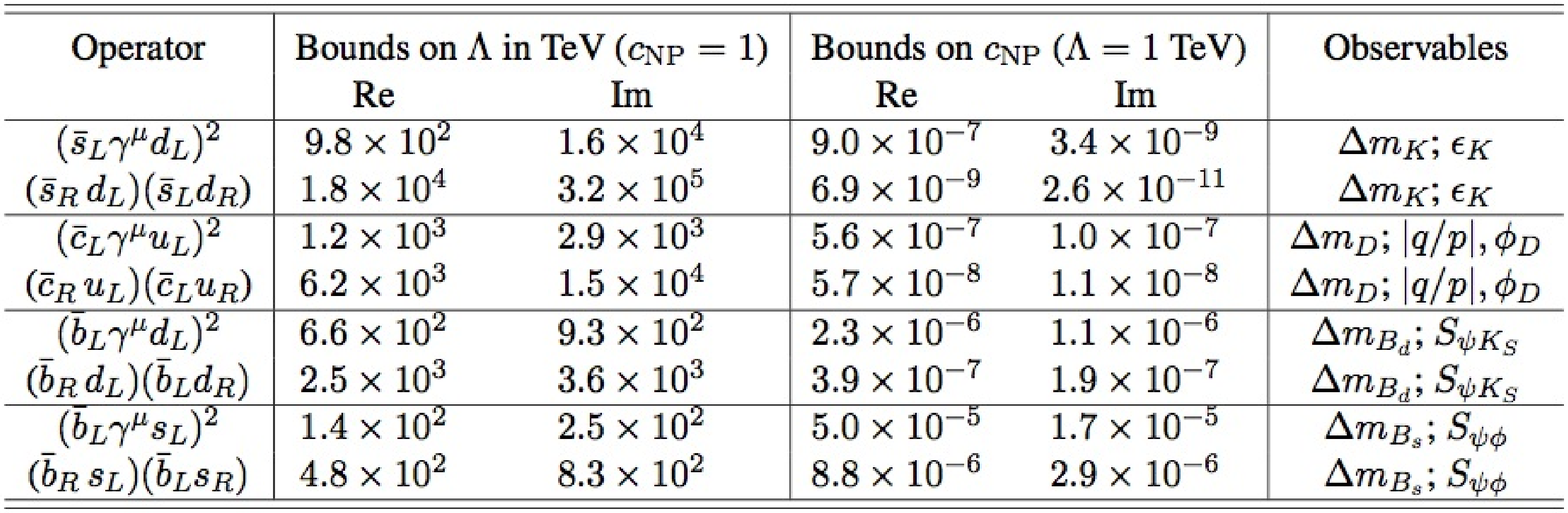}
\caption{{\small Bounds on representative, dimension-six, $\Delta$F=2 operators, assuming an effective coupling $c_{NP}/\Lambda^2$. The bounds are quoted on $\Lambda$, setting $|c_{NP}|$ = 1, or on $c_{NP}$, setting $\Lambda$=1 TeV. The right column indicates the main observables used to derive these bounds.}}
\label{fig:bounds}
\end{center}
\end{figure}

A recent analysis shows that the situation embodied in the above equation, with $c_{NP}\approx 1$ leads to very large values of the cutoff, see Fig. \ref{fig:bounds} taken from  \cite{isidorilast}. Evidently NP at the TeV scale cannot be coupled to flavour in a generic way. Some extended GIM mechanism is required. 

Many insights and many interesting papers have gone in this subject over the last years, see again \cite{isidorilast} for references.  

\section{Minimal Flavor Violation}

In the Standard Theory there is a large, global, group, ${\cal G}$, associated to flavor, which commutes with the gauge group $SU(3)\otimes SU(2)\otimes U(1)$.
With three generations of two weak left-handed doublets of quark and leptons, $Q_L$, $L_L$, and three right-handed singlets, $U_R$, $D_R$, $E_R$, one has ${\cal G}=U(3)^5$.

In the Standard Theory, the symmetry group ${\cal G}$ is broken by the Yukawa couplings of the above multiplets to the Higgs doublet, which, neglecting neutrino masses, can be written schematically as:
\be
{\cal L}_Y={\bar Q}_L {\it Y}_D H D_R +{\bar Q}_L {\it Y}_U {\tilde H} U_R+{\bar L}_L {\it Y}_E H E_R 
\label{yukawac} 
\ee
where H is the Higgs doublet, ${\tilde H}$ the charge conjugate field and the $Y$s adimensional coupling constants arranged each in three by three complex matrices. After the Higgs field takes a vacuum expectation value, the Yukawa lagrangian gives rise to the fermion mass matrices according to:
\bea 
&&  {\cal L}_{mass}= {\bar D}_L {\it M}_D D_R+{\bar U}_L {\it M}_U U_R +{\bar E}_L {\it M}_E E_R \nonumber \\
&& {\it M}_D=<H_0>{\it Y}_D,\;{\rm etc.}
\label{fermmass}
\eea

The lagrangian (\ref{yukawac}) would be invariant under ${\cal G}$ if we would subject the $Y$s to the same transformations of the fields. In reality, this is not true since the $Y$s are numerical constants, but the trick, introduced long ago to describe the breaking of a symmetry\footnote{the term {\it spurion} was coined in this context for the symmetry breaking parameters analogous to the $Y$s.} and considered in the flavor context first by Georgi and Chivukula \cite{georgichiv}, allows for an efficient book keeping of the effects of (\ref{yukawac}). 

The principle of Minimal Flavor Violation (MFV) can now be stated as follows \cite{dambrosioetal}: Yukawa couplings are the only source of flavor symmetry violation, for the old and for the hypothetical new physics.

The {\it spurion} picture helps to understand how this may work. NP effects will add non renormalizable, dimension six operators to the effective lagrangian, as in eq. (\ref{np_efflagr}). These terms, however, must contain appropriate powers of the $Y$s, so as to make these operators invariant under ${\cal G}$, if we subject the $Y$s and the fields to the appropriate transformations. In this way, powers of quark masses and CKM couplings, contained in the $Y$s, will appear in the effective operators ${\cal O}_i$, so as to mimick the suppressions embodied by the GIM mechanism in, e.g., eq. (\ref{deltam}).
How do we understand that the Yukawa couplings may appear in the NP sector? 

In the original version of Georgi and Chivukula, the $Y$s were related to the {\it preon mass terms}. So, they would affect quark and lepton physics much in the same way that chiral symmetry breaking parameters, i.e. the quark masses, affect hadron physics. In SUSY, MFV amounts to say that ${\cal G}$ breaking appears only once, in the Yukawa couplings of the supermultiplets, presumably at the Grand-Unification scale (i.e. the soft-breaking terms feel the breaking of G only via the Yukawa couplings).

\begin{figure}[ht]
\begin{center}
\includegraphics[scale=.5]{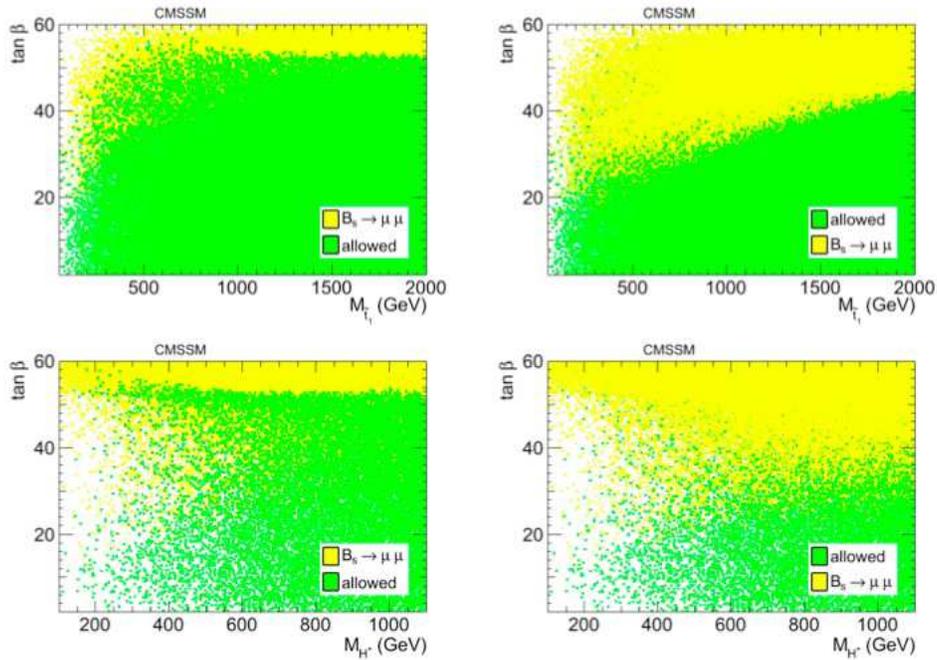}
\caption{{\small Constraints from $BR(B_s \to \mu^+\mu^-)$ in the CMSSM plane ($M_{\tilde t}, \tan\beta$) in the upper panel and ($M_{H^\pm} , \tan \beta$) in the lower panel, with the allowed points displayed in the foreground in the left and in the background in the right.}}
\label{bstomumuSUSY}
\end{center}
\end{figure}

An example of MFV is given by the Constrained Minimal SuperSymmetric Model. 
For illustration, I reported in Fig. \ref{bstomumuSUSY} the limits one finds \cite{nazila}in CMSSM from the observed branching ratio of $B_s\to \mu^+\mu^-$ for the mass of the scalar top or the charged Higgs boson predicted by MSSM. One sees that the limits derived from this particular FCNC process are compatible with a relatively low energy scale for New Physics (for more details and recent comparison of MFV with data see \cite{isidoristraub}). 

Alternatives to MFV are being studied in models where quark and lepton masses are obtained from the mixing of the elementary fermions with composite fermions at higher energy \cite{continopomarol}. For a recent  review, see \cite{barbierietal}.

\paragraph{Are Yukawa couplings the VEVs of new fields?} Promoting the spurion to a real field, whose vacuum expectation value gives rise to the Yukawa couplings is an idea pioneered by Froggat and Nielsen \cite{frogniels} who associated these fields to a $U(1)$ symmetry.
In the same spirit, the Yukawa couplings could be determined by a variational principle, i.e. by the minimum of a new hidden potential with the symmetry of the flavor group ${\cal G}=U(5)^5$, or variations thereof, to include neutrino masses.

Interesting applications to neutrino masses and mixing have been recently reported by B. Gavela and coll. ~\cite{Alonso:2011yg,Alonso:2012fy}
(see also Gavela's talk at this Conference). 

Long ago, Nicola Cabibbo speculated that the value of the weak interaction angle could be derived by a minimum principle obeyed by the chiral symmetry breaking. It is interesting to speculate that the idea can be revived within the flavor group ${\cal G}$. The methods then derived \cite{michel,cabmaia} 
can be applied to find indications for quark and neutrino Yukawa couplings.

\section{Conclusions}
Checking selection rules has been an effective way to guess new physics at higher energy.

The suppression of $\Delta$S=1, 2 neutral current processes led to the charm quark and to CP violation with the third generation.
Several processes, besides those considered here, may give useful information and are actively searched:
\bea
&& K^+\to \pi^+ \nu \bar \nu;\; K_L\to \pi^0 \nu \bar \nu; \;
B_d\to \mu^+\mu^-; \;b\to s \gamma;\\
&& \mu \to e\gamma; \;\mu+N\to e+\cdots
\eea

The effectiveness of the GIM mechanism to describe the observed FCNC suppression gives already important restrictions on what may be the physics beyond the Standard Theory. 
And it gives insights on the nature of the Yukawa couplings and the breaking of the global flavor symmetry. 

Hopes are not lost to find effects of New Phyics at accessible energies by detecting small deviations from ST predictions in high precision experiments.

\section*{Acknowledgments}

Hospitality at the CERN Theory Group, where this lecture was prepared, is gratefully acknowledged. I am grateful to Belen Gavela,  Gino Isidori, John Iliopoulos,  Antonello Polosa and Veronica Riquer for illuminating discussions.

\end{document}